\documentclass{iaus}
\usepackage{graphicx}

\title[SMC in space and time] %% give here short title %%
{SMC in space and time: a project to study the evolution of the
prototype interacting late-type dwarf galaxy}

\author[M. Tosi et al.]   %% give here short author list %%
{M.Tosi$^1$
J.Gallagher$^2$, E.Sabbi$^3$, K.Glatt$^{4,5}$, E.K.Grebel$^4$, 
C.Christian$^3$, M.Cignoni$^{6,1}$, G.Clementini$^1$, A.Cole$^7$, 
G.Da Costa$^8$, D.Harbeck$^2$, M.Marconi$^9$, M.Meixner$^3$, 
A.Nota$^3$, M.Sirianni$^3$ \and T.Smecker-Hane$^{10}$}

\affiliation{$^1$INAF - Osservatorio Astronomico di Bologna \\ Via Ranzani 1,
I-40127, Bologna, Italy \\ email: {\tt monica.tosi@oabo.inaf.it}\\[\affilskip]
$^2$University of Wisconsin, Madison, WI, USA \\[\affilskip]
$^3$STScI, Baltimore, MD, USA \\[\affilskip]
$^4$Heidelberg University, Heidelberg, D \\[\affilskip]
$^5$Basel University, Basel, CH \\[\affilskip]
$^6$Bologna University, Bologna, I \\[\affilskip]
$^7$University of Tasmania, AU  \\[\affilskip]
$^8$Research School of Astronomy \& Astrophysics, ANU, AU \\[\affilskip]
$^9$INAF - OA Capodimonte, Napoli, I \\[\affilskip]
$^{10}$University of California, Irvine, CA, USA  
}

\pubyear{2008}
\volume{255}  %% insert here IAU Symposium No.
\pagerange{}
\jname{Low-Metallicity Star Formation: From the First Stars to Dwarf Galaxies}
\editors{L.K. Hunt, S. Madden \& R. Schneider, eds.}
\begin{document}

\maketitle

\begin{abstract}
We introduce the {\it SMC in space and time}, a large coordinated space and 
ground-based program to study star formation processes and history,
as well as variable stars, structure, kinematics and chemical evolution of
the whole SMC.
Here, we present the Colour-Magnitude Diagrams(CMDs) resulting from HST/ACS 
photometry, aimed at deriving the star formation history (SFH) in six fields of 
the SMC. The fields are located in the central regions, in the stellar halo, and
in the wing toward the 
LMC. The CMDs are very deep, well beyond the oldest Main Sequence Turn-Off, and 
will allow us to derive the SFH over the entire Hubble time.
%% add here a maximum of 10 keywords, to be taken form the file <Keywords.txt>
\end{abstract}

\firstsection % if your document starts with a section,
              % remove some space above using this command.
\section{Introduction}
The Small Magellanic Cloud (SMC) is the closest late-type dwarf and has many
properties similar to those of the vast majority of this common class of 
galaxies. Its current metallicity (Z$\simeq$0.004 in mass fraction, as derived 
from HII regions and young stars) is typical of dwarf irregular and Blue 
Compact Dwarf (BCD) galaxies, the least evolved systems, hence the most similar to primeval 
galaxies. Its mass (between 1 and 5 $\times 10^9 M_{\odot}$, e.g. 
\cite[Kallivayalil et al. 2006]{K06} and references therein) is at
the upper limit of the range of masses typical of late-type dwarfs. These
characteristics, combined with its proximity, make the SMC the natural 
benchmark to study the evolution of late-type dwarf galaxies. Moreover,
its membership to a triple system allows detailed studies of
interaction-driven modulations of the star formation activity.

A wealth of data on the SMC are available in the literature, although not as 
much as for its bigger companion, the LMC. Yet, much more are needed for a 
better
understanding of how the SMC has formed and evolved. We have thus embarked on a
long-term project to study the evolution of the SMC in space and time. 
Our project plans to
exploit the high performances in depth, resolving power or large field of view
of current and forthcoming, space and ground based, telescopes, such as HST, 
VLT, Spitzer, SALT and VST.

\section{The SMC in Space and Time}

Primary goals of our long-term project are the derivation of the star 
formation
history (SFH) in the whole SMC from deep and accurate photometry  and of 
stellar chemical abundances from high and intermediate resolution spectra. 
These data will allow us to infer the
age-metallicity relation (or lack thereof) of stars resolved in different
regions of the galaxy and to better constrain numerical models for the chemical
evolution of the various SMC regions as well as for the galaxy as a whole. 
Since the SFH and the age-metallicity 
relation are key parameters in chemical evolution modeling, these new models 
will be
of unprecedented accuracy for an external galaxy, reaching the level of
reliability currently attained only for the solar neighbourhood.

Part of the project is devoted to the study of the SMC variable stars of all
types, to classify them and use the unique aspects of variability to
get their physical properties. We will study the
spatial distribution of the various types of variables and this will provide
unique information on the space and time confinement of the formation of their
parent stellar populations. Standard candles such as the RR Lyraes will also
provide information on the 3D structure of the galaxy and on its reddening
distribution.

The SFH will be derived from Colour-Magnitude diagrams (CMD) using the synthetic
CMD technique. This kind of study has already been performed by other authors
(e.g. \cite[Dolphin 2001]{Dolp01}, \cite[Harris \& Zaritsky 2004]{HZ04},
\cite[Chiosi et al. 2006]{C06}, \cite[Noel et al. 2007]{N07}). Our plan, however,
is to have CMDs several magnitudes fainter than the oldest main-sequence (MS)
turn-off (TO) for the entire galaxy, including its halo and the wing in the
direction of the LMC, allowing us to infer for the first time the SFH of the
whole SMC over the entire Hubble time.

\begin{figure}[b]
% \vspace*{-2.0 cm}
\begin{center}
 \includegraphics[width=13cm]{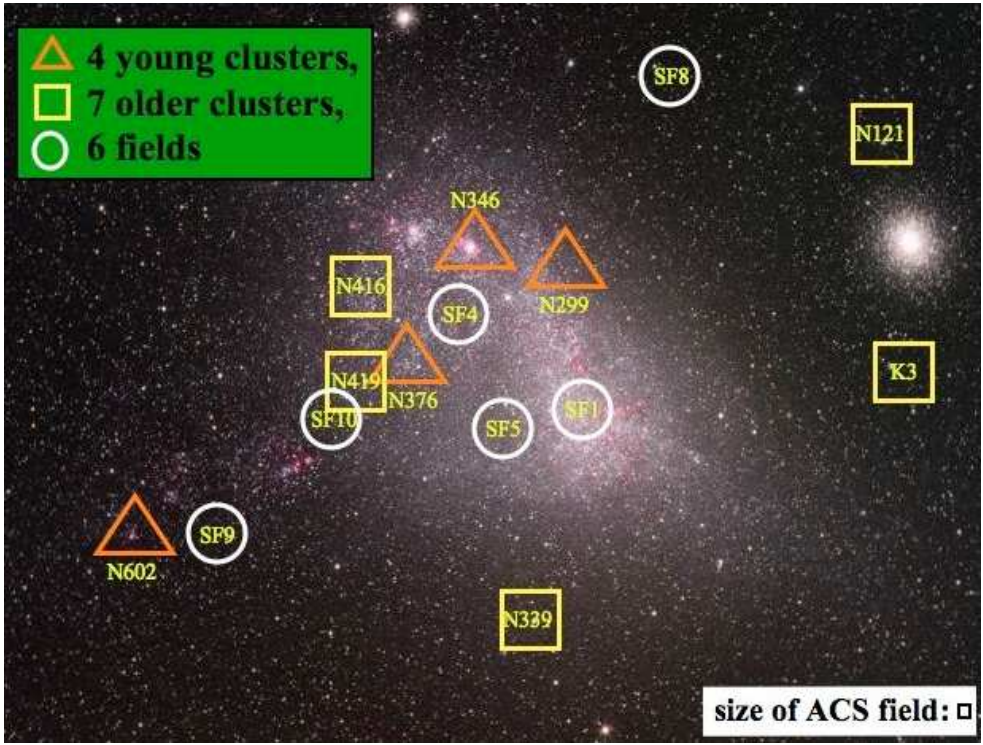} 
% \vspace*{-1.0 cm}
 \caption{Location of our HST/ACS targets, overimposed on an SMC image (courtesy
 St\'ephane Guisard). The young clusters are indicated by (orange) triangles, the older
 clusters by (yellow) squares, and the fields by the white circles. The name of
 the target is shown, with N standing for NGC and K for Kron. Lindsay 1 and
 Lindsay 38 fall out of the shown sky area. The
 actual size of the ACS field is shown in the bottom-right corner. }
   \label{fig1}
\end{center}
\end{figure}

Time has already been awarded to this project on HST (PIs A. Nota and J.
Gallagher) and on VLT (PI E.K. Grebel) and is guaranteed on VST (PI V. Ripepi).
HST/ACS photometry was acquired in Cycle 13 for 4 young star clusters, 7 older
clusters and 6 fields. Fig.\,\ref{fig1} shows the location of our HST/ACS 
targets (except for the two outermost old clusters). Results from these
data have already been published on the young clusters NGC~346 
(\cite[Nota et al. 2006]{N06} and
\cite[Sabbi et al. 2007]{S07}) and NGC~602 (\cite[Carlson et al. 2007]{C07}) 
and on the seven old clusters (\cite[Glatt et al. 2008a]{G08a} for NGC~121, and 
\cite[Glatt et al. 2008b]{G08b} for NGC~339, NGC~416, NGC~419, 
Lindsay 1, Lindsay 38 and Kron 3). 
More details on the young clusters are presented by Sabbi et al. in this
volume, while the derivation of the SFH in the region of NGC~602 is described by 
\cite[Cignoni et al. (2008)]{C08}. 

\section{The HST/ACS fields}

The six SMC fields have been observed in Cycle 13 (GO 10396, PI J. 
Gallagher) with the Wide Field Channel of the 
HST/ACS in the F555W (V) and F814 (I) bands for a total of 12 orbits. The target
fields (indicated by white circles in Fig.~\ref{fig1}) have been chosen to
maximize possible stellar population differences between different SMC regions. Three
fields are in the central region: one (SF4) close to the barycenter of the young 
population, one (SF1) close to the barycenter of the old population, and one
(SF5)
in an intermediate zone. Two fields are located in the wing, the SMC extension
towards the LMC: one (SF9) in the wing outer part, and one (SF10) in its inner
part. The last field (SF8) is in the opposite side, in what can be considered
the SMC halo.

The observations were performed with the ACS/WFC, following a standard dithering 
pattern to improve PSF sampling, allow for hot pixel and cosmic ray removal 
and fill the gap between the two WFC detectors. The photometric analysis has been performed
independently with two packages suited for PSF fitting in crowded fields:
Stetson's Daophot and Anderson's imgxy-WFC.01x10.
Extensive artifical star tests have been performed on the images to assess
photometric errors, incompleteness and blending factors.

\begin{figure}[b]
% \vspace*{-2.0 cm}
\begin{center}
 \includegraphics[width=13cm]{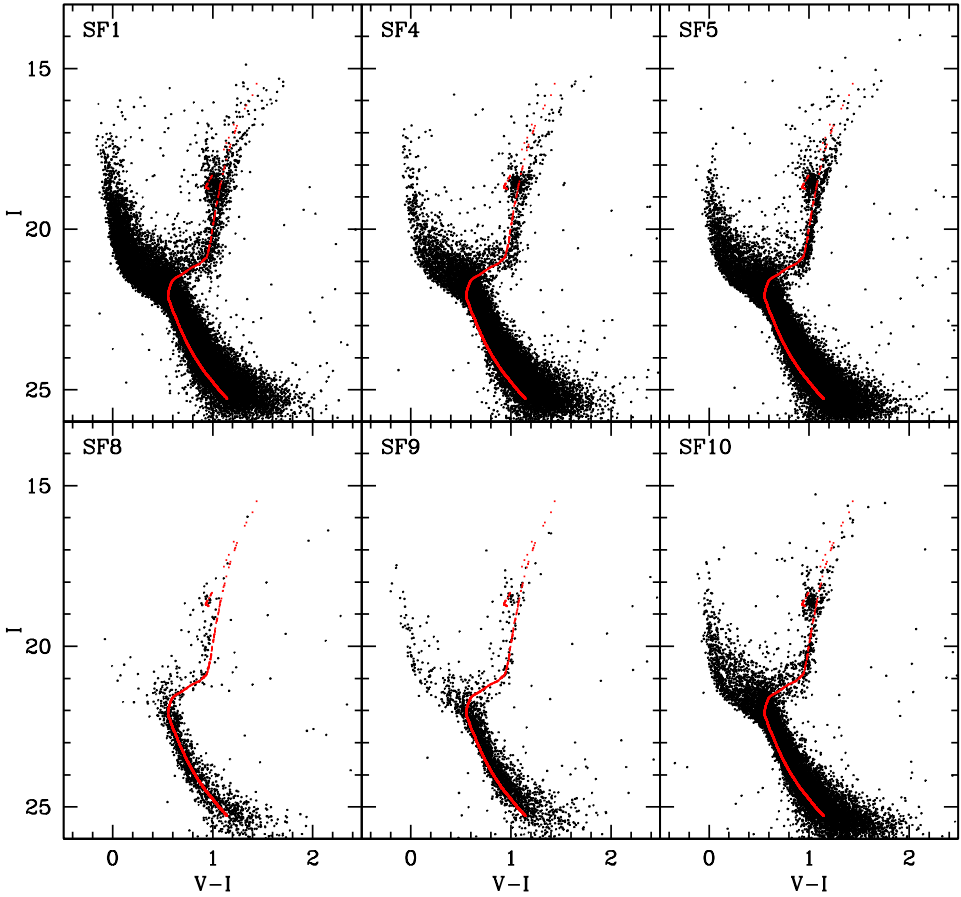} 
% \vspace*{-1.0 cm}
 \caption{CMDs of the six SMC fields observed with HST/ACS. The three top panels
refer to the central regions: SF1, with 28000 measured stars, SF4, with
16000 stars, and SF5, with 25300 stars. The bottom-left panel refer to the
halo field, SF8, with 1550 stars. The central bottom panel shows the CMD of
the outer wing field SF9, with 2440 stars, while the CMD of the inner wing
field, SF10, with 13700 stars, is in the bottom-right panel. The 7 Gyr
isochrone with Z=0.001,
interpolated by \cite[Angeretti et al. (2007)]{An07} from the Padova
stellar evolution models (\cite[Fagotto et al. 1994a]{Fa94a} and 
\cite[Fagotto et al. 1994b]{Fa94b}) is also shown in 
all panels for reference. The assumed intrinsic distance modulus and reddening
are (m-M)$_0$=18.9 and E(B-V)=0.08.}
   \label{cmd}
\end{center}
\end{figure}

The CMDs resulting from the application of Anderson's photometric package are 
shown in Fig.\ref{cmd}, where a 7 Gyr isochrone with metallicity Z=0.001 (from
\cite[Angeretti et al. 2007]{An07}) is 
also plotted for reference. As in all resolved galaxies, there is a clear 
stellar density gradient from the SMC center to the periphery, the field
containing the largest number of stars (25300) being SF5.
All regions turn out to contain old  stellar
populations whose evolutionary phases are visible in the CMDs: 
main-sequence (MS), subgiant branch (SGB), red giant branch (RGB), clump and 
asymptotic giant branch (AGB). 
The age of these old populations appears to be of several Gyr, mostly
around 7 Gyr. 

All fields, but the halo one, also show the blue plume typical of
late-type dwarf galaxies, populated by high and intermediate mass stars in the 
main-sequence phase or at the blue edge of the blue loops (corresponding to the
central He-burning phase). 
It is interesting to notice the contrast between the outer wing, where
in spite of a relatively low number of measured stars the blue plume is well
populated with young stars, and the much older halo in the SF8 field.

An interesting feature of the six CMDs is the apparent homogeneity of the old 
populations: in all panels of Fig.\ref{cmd} a) the old SGBs  have 
roughly the same magnitude and the old MS TOs have roughly 
the same colour, and b) the clumps have roughly the same magnitude and colour. 
This circumstance suggests that no large differences in age 
and metallicity exist among the SMC old stars, irrespectively of their
spatial location. By inspecting the CMDs in more detail (for instance with the
aid of the 7 Gyr isochrone), we do see that the stars in the halo field 
(SF8) are bluer than
the others, presumably because they are metal poorer and/or less reddened,
whilst those at the barycenter of the SMC young population
(SF4) are the reddest of all, probably because they are metal richer or more reddened.
Also the widths of the various evolutionary phases appear somewhat different
from one field to the other, which could be due either to actual age or
metallicity spreads or to differences in the distribution of stars in distance
within the SMC. At any rate,
the CMDs are too similar to each other in the phases relative to old stars to
allow for macroscopic differences in the early evolution of the six fields. 
The six regions seem to share a relatively late onset of the bulk of star
formation activity, not much earlier than 7 Gyr ago. This result is in 
agreement with the findings 
by \cite[Dolphin et al. (2001)]{Dolp01} and \cite[Noel et al. (2007)]{N07} for
other SMC regions, and at variance 
with \cite[Harris \& Zaritsky (2004)]{HZ04} conclusions, based, however,
on shallower photometry not reaching the old MS TO.

To infer the details of the SFH in the six regions, fully exploiting all the photometric
information, we will apply the synthetic CMD method taking into account
photometric errors, incompleteness and crowding effects as estimated with the
artificial star tests.
For a better assessment of the uncertainties involved in the SFH derivation, we
will apply three different and independent approaches of the synthetic CMD
method: Cignoni's (see e.g. \cite[Cignoni et al. 2006]{Ci06}), Cole's (see e.g.
\cite[Cole et al. 2007]{Co07}) and Tosi's (see e.g. \cite[Tosi et al.
1991]{To91} and \cite[Angeretti et al. 2005]{An05}).
The CMDs presented here already allow to forecast interesting results. 
When the VST will be operating and we will cover the whole SMC, wing included,
with B, V, I photometry reaching several magnitudes below the oldest MS
TO, we will be able to assess precisely how the star formation activity
has evolved both in space and in time.

\smallskip\noindent{\bf Acknowledgements}
This work has been partially supported through NASA-HST funding to JSG, PRIN-INAF-2005,
and Swiss NSF 200020-113697.

\end{document}